\documentclass{article}

\usepackage{PRIMEarxiv}

\usepackage[utf8]{inputenc} 
\usepackage[T1]{fontenc}    
\usepackage{hyperref}       
\usepackage{url}            
\usepackage{booktabs}       
\usepackage{amsfonts}
\usepackage{amsthm}
\usepackage{nicefrac}       
\usepackage{microtype}      
\usepackage{lipsum}
\usepackage{fancyhdr}       
\usepackage{graphicx}       
\usepackage{cite}
\usepackage{amsmath,amssymb,amsfonts, mathabx}
\usepackage{algorithmic}
\usepackage{graphicx}
\usepackage{textcomp}
\usepackage{xcolor}
\usepackage{cite}
\usepackage{float}
\usepackage{mathtools}
\usepackage{xspace}
\usepackage{xcolor}
\usepackage{graphicx}
\usepackage{appendix}

\usepackage{algorithm}
\usepackage{algorithmic}
\floatname{algorithm}{Procedure}
\newcommand{\Procedure}[2]{\STATE \textbf{Procedure} #1(#2)}
\newcommand{\BeginProcedure}{\begin{ALC@g}}
\newcommand{\EndProcedure}{\end{ALC@g} \STATE \textbf{end procedure}}

\makeatletter
\newcommand{\removelatexerror}{\let\@latex@error\@gobble}
\makeatother

\newtheorem{definition}{Definition}
\newtheorem{theorem}{Theorem}
\newtheorem{lemma}{Lemma}
\newtheorem{corollary}{Corollary}

\newcommand{\ouralg}{{\textsf{GTUA}}\xspace}
\newcommand{\laalg}{{\textsf{LA}}\xspace}
\newcommand{\gbs}{{\textsf{GBS}}\xspace}

\title{Behavior-Aware Efficient Detection of Malicious EVs in V2G Systems}

\author{
  Ruixiang Wu, Xudong Wang \\
  School of Data Science\\
  The Chinese University of Hong Kong, Shenzhen \\
  Shenzhen, Guangdong, China\\
  \texttt{\{ruixiangwu, xudongwang\}@link.cuhk.edu.cn} \\
   \And
  Tongxin Li\\
  School of Data Scienc \\
  The Chinese University of Hong Kong, Shenzhen \\
  Shenzhen, Guangdong, China\\
  \texttt{litongxin@cuhk.edu.cn} \\
}

\begin{document}
\maketitle

\begin{abstract}
With the rapid development of electric vehicles (EVs) and vehicle-to-grid (V2G) technology, detecting malicious EV drivers is becoming increasingly important for the reliability and efficiency of smart grids. To address this challenge, machine learning (ML) algorithms are employed to predict user behavior and identify patterns of non-cooperation. However, the ML predictions are often untrusted, which can significantly degrade the performance of existing algorithms. In this paper, we propose a safety-enabled group testing scheme, \ouralg, which combines the efficiency of probabilistic group testing with ML predictions and the robustness of combinatorial group testing. We prove that \ouralg is $O(d)$-consistent and $O(d\log n)$-robust, striking a near-optimal trade-off. Experiments on synthetic data and case studies based on \textsc{ACN-Data}, a real-world EV charging dataset validate the efficacy of \ouralg for efficiently detecting malicious users in V2G systems. Our findings contribute to the growing field of algorithms with predictions and provide insights for incorporating distributional ML advice into algorithmic decision-making in energy and transportation-related systems.
\end{abstract}


\section{Introduction}
Recently, with the rising demand for green technology in modern cities, the electric vehicle (EV) industry is experiencing rapid development in smart cities~\cite{abdullahi2024hybrid}. Integrating EVs into smart cities presents a significant potential to cut greenhouse gas emissions~\cite{10571532}. One promising approach for this integration is Vehicle-to-Grid (V2G), which allows bidirectional power exchange between EVs and the smart grid~\cite{7744655}. This enables EVs to charge from the smart grid during off-peak hours and supply electricity to the grid during peak periods, further enhancing the flexibility of the power systems, and boosting the implementation of virtual power plant in smart cities.~\cite{9855066, 10415173}. Furthermore, V2G can provides significant economic benefits for EV owners by lowering ownership costs~\cite{RePEc:inm:ormnsc:v:66:y:2020:i:9:p:4152-4172}. Collectively, these benefits position V2G as a highly promising technology for the future.

While the V2G system provides significant flexibility to the smart grid, it also demands considerable flexibility from its users~\cite{QIN2023100291}, and its reliability and efficiency are closely related to users' behaviors~\cite{shin2024smart}. The issues will arise when V2G users are either malicious or lack flexibility. These users might ignore the committed sojourn times and drive their EVs away during peak hours after getting charged during off-peak hours, preventing the V2G system from accessing the stored electricity. 

Although installing meters at each charger allows for individual monitoring, the large-scale data generated introduces substantial financial and logistical burdens, and transmitting such data exposes users to privacy risks~\cite{10.1007/978-3-031-51674-0_7, 10083184}. Additionally, the cost for a residential electric meter ranges from $\$60$ to $\$250$~\cite{evvr2024smartmeter}, this approach may not be practical for detecting malicious users in a large-scale power grid due to the high expense of equipping each user with sensors. Thus, a more efficient technique is crucial for reducing data expenses and enhancing privacy while effectively identifying malicious users.

\subsection{Previous Work}
Previous work has explored economic incentive approaches to deter malicious behavior in V2G systems. For instance, Vickrey-Clarke-Groves (VCG) auction-based algorithms have been proposed to encourage the participation of EVs in V2G networks by using penalty payments to promote honest reporting and discourage malicious activity~\cite{10054045, 7249160}. Despite their advantages, VCG algorithms have notable limitations. They are susceptible to false-name attacks, as highlighted in~\cite{gafni2021vcgfalsenameattacksbayesian}. Furthermore, the reliance on exact optimization renders them computationally infeasible in complex scenarios, and also, the VCG mechanism often fails to adequately address strategic behaviors that deviate from simple truth-telling~\cite{dobzinski2011limitations}.

To address these challenges, an effective approach is to use group testing algorithms to directly detect malicious users, offering a more robust alternative to the relatively ``soft" method provided by VCG mechanisms. Group testing algorithms aim to reduce the number of tests by grouping individuals collectively~\cite{du2000combinatorial,comp}. More recently, group testing algorithms have been applied in various areas of smart grid like fault identification and system reliability assessment. The Adaptive Binary Splitting Inspection (ABSI) algorithm~\cite{8409488} employs binary search techniques to efficiently identify malicious users in large-scale smart grid environments. In 2020, Xia et al.~\cite{Xia2020GTHI} proposed an inspection algorithm based on group testing to detect malicious users. This approach adaptively switches between individual and group testing strategies to achieve higher efficiency. However, the aforementioned group testing algorithms did not consider user history behavior, which is easily obtainable in a smart grid environment. In this paper, our proposed algorithm \ouralg fully incorporates user history behavior, thereby improving the efficiency of the testing process.

\subsection{Contributions}
\begin{figure}[!t]
    \centering
\includegraphics[width = 0.7\textwidth]{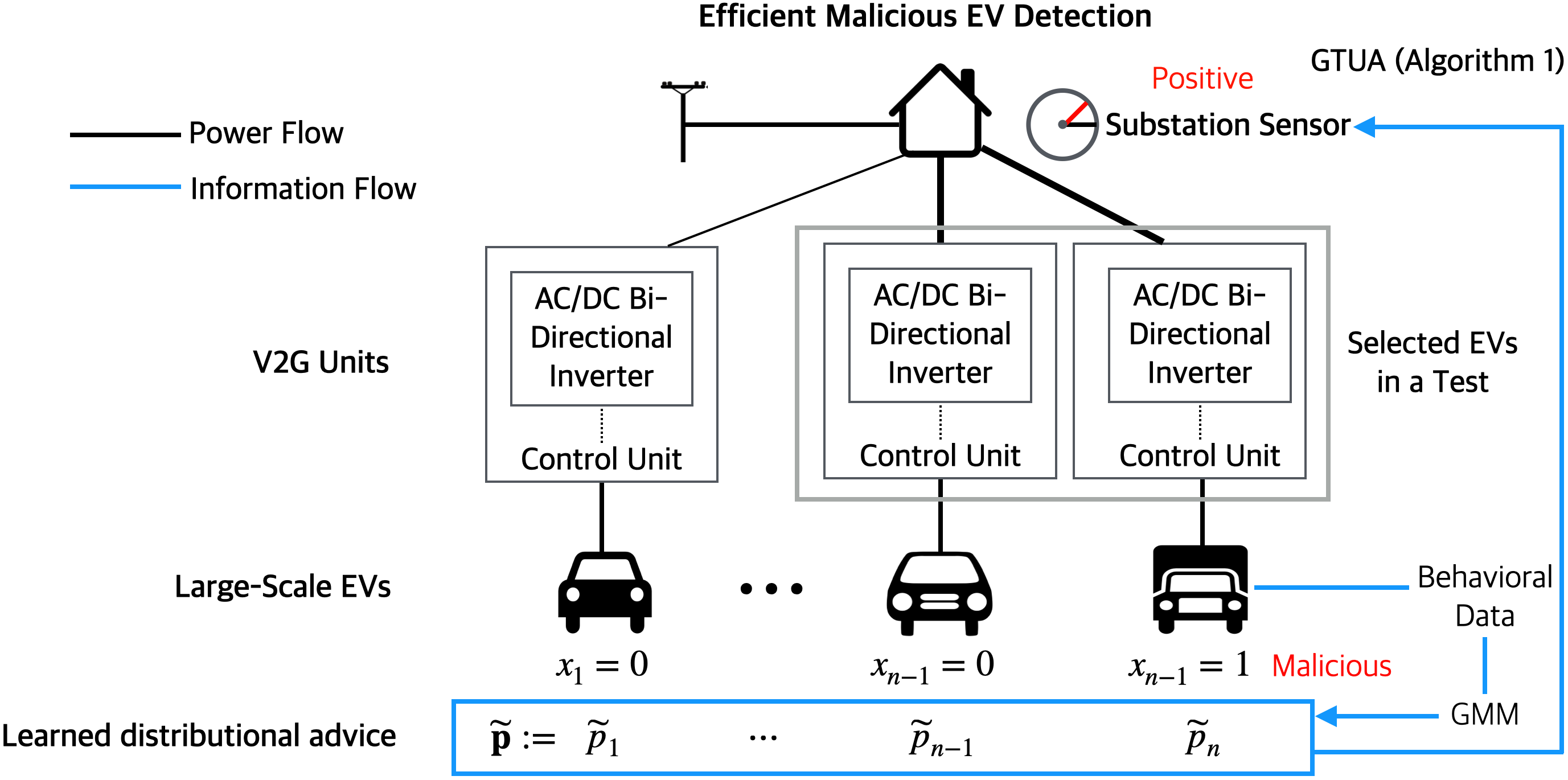}
\caption{Illustration of Malicious User Detection in a V2G System (see the model defined in Section~\ref{sec:model}). A V2G manager is equipped with a sensor to detect malicious EVs from a subset of EVs specified by an algorithm. Machine learning models learned from grid data generate distributional advice $\widetilde{p}$~(denoting the probabilities of behaving maliciously). An algorithm, denoted by~\ouralg is presented in Section~\ref{sec:GT+MLPred} for efficient malicious EV detection.}
    \label{fig:maliciousdemo}
\end{figure}

Our main contributions are: (1) We reformulate the detection task as a novel \textit{group testing} problem that utilizes untrusted machine-learned \textit{distributional advice} to detect malicious users in a V2G system, as illustrated in Figure~\ref{fig:maliciousdemo}; (2) Introduce \ouralg, a safety-enabled testing method (Algorithm~\ref{alg:pseudocode}) combining combinatorial and probabilistic strategies; (3) We provide theoretical bounds on the expected $\kappa$-regret (Theorem~\ref{thm:regret}), demonstrating robust utilization of predicted statistics; (4) We establish that our method achieves \textit{$O(d)$-consistency} and \textit{$O(d\log n)$-robustness} (Corollary~\ref{coro:tradeoff}), where $d$ is the number of malicious drivers and $n$ the total drivers, showing near-optimal trade-offs for group testing algorithms.

\section{Problem Formulation} \label{sec:model}
Consider a V2G system consisting of a large fleet of EV chargers. To detect malicious EVs, an efficient method is to remotely connect a carefully selected subset of EVs and monitor the substation power injection from the remaining chargers to see if it meets the expected setpoints (i.e., checking if the connected EVs behave normally), requiring only a single sensor installed at the substation.  If the substation power injection from the fixed subset is far from the expected nominal value, it indicates that the presence of malicious EV(s). In summary, for a detection within a subset of EVs, if at least one EV behaves maliciously, the substation power injection will be abnormal (marked as $1$); otherwise, it will be normal (marked as $0$).
Mathematically, each run of this detecting process can be represented by an ``OR" operation, as described in the following detection model.

\textbf{Malicious EV Detection Model.} Consider a set $\mathcal{N}$ of $n$ EV drivers, wherein each driver $i$ is malicious with some probability $p_i\in [0,1]$ independently. Let $[n] \coloneqq \{1,\ldots,n\}$.
The states of the $n$ drivers are represented by an $n$-length vector $\mathbf{X} = \left(X_i:i\in [n]\right)$ such that $X_i=1$ (or $X_i=0$) implies that the $i$-th driver is malicious (or not). Our goal is to asymptotically detect malicious drivers by conducting a minimal number of tests. Each test selects a subset of drivers and performs an ``OR" operation, which is defined as follows: The test outcome is positive if the subset contains at least one malicious EV (i.e., there exists at least one $X_i = 1$ within the subset); otherwise, it is negative. 

\subsection{Untrusted ML Predictions}
With data available in many real-world scenarios, prior statistics of the probabilities $\mathbf{p}\coloneq (p_1,\ldots,p_{n})$ can often be predicted via black-box ML models. The problem considered in this paper, which sets it apart from the classic probabilistic group testing models~\cite{comp,aldridge2019group}, involves incorporating a set of predictions $\widetilde{\mathbf{p}} = (\widetilde{p}_1,\ldots,\widetilde{p}_n)$ of $\mathbf{p}$. These predictions are assumed to be provided by an ML agency whose reliability is not fully trusted. The potential for error in $\widetilde{\mathbf{p}}$ arises from various factors, such as the inherent variability in learning algorithms~\cite{recht2019tour} or challenges in generalizing beyond the distribution of the training data~\cite{wenzel2022assaying}, etc.
Our model takes into account these uncertainties in $\widetilde{\mathbf{p}}$.

Let $\mathcal{P}\coloneqq \left\{\mathbf{p}:0\leq p_i\leq 1, i\in [n]\right\}$ be a set consisting of all $\mathbf{p}$'s. Denote by $d\coloneqq \sum_{i=1}^{n}p_i$. According to Byzantine agreement~\cite{lamport1982the}, for a system to achieve reliable consensus in the presence of malicious nodes, the number of faulty nodes must satisfy the condition above. Accordingly, we focus on the sub-linear regime $d=o(n)$ aligning with the standard sparsity assumption in the classic group testing model. The \textit{distributional advice} given by an untrusted ML model is denoted by $\widetilde{\mathbf{p}} \coloneqq \left(\widetilde{p}_i:i\in [n]\right)\in\mathcal{P}$ such that $\sum_{i=1}^{n}\widetilde{p}_i=d$.\footnote{It is known that standard concentration inequalities imply that with high probability, the
actual number of malicious users can be bounded by  $\delta d$ for some $\delta>1$~\cite{comp,priorstats}. Thus, WLOG, we assume the predictions are normalized such that $\sum_{i=1}^{n} \widetilde{p}_i=d$.} To characterize the prediction error, we use the following accumulated pseudo KL-divergence  (logarithm base $e$) between $\mathbf{p}$ and $\widetilde{\mathbf{p}}$:
\begin{equation}
\label{eq:sum_KL}
D_{\mathrm{KL}}\left(p,\widetilde{p}\right) \coloneqq \sum_{i=1}^{n} \left(p_i\log\frac{p_i}{\widetilde{p}_i} \right).
\end{equation}

In this work, we focus on designing \textit{adaptive} group testing algorithms, with access to untrusted predictions $\widetilde{p}$, i.e., the next test can be determined after observing previous test outcomes. We revisit the classic adaptive probabilistic group testing problem within the framework of online optimization. This perspective allows us to assess our adaptive group testing algorithm against established performance benchmarks commonly used in online optimization contexts. The specifics of the benchmark used in this paper are detailed next.

\subsection{Performance Benchmarks}
Our algorithm is evaluated based on a novel benchmark known as the \textit{$\kappa$-regret}. This benchmark hinges on an external parameter $\kappa$ that measures the performance or ``tightness'' of the current best algorithms in probabilistic group testing. This parameter is defined as the \textit{tightness gap} below. Consider any deterministic group testing algorithm, denoted as $\textsf{ALG}$. Let $\overline{T}(\textsf{ALG};\mathbf{p})$ represent the best-known upper bounds on the expected number of tests needed by $\textsf{ALG}$. We then define $\underline{T}^*(\mathbf{p})$ as the minimum expected number of tests required, considering the same prior information $\mathbf{p}$. 

\begin{definition}[Tightness Gap]\label{def:regret}
Given a deterministic group testing algorithm $\textsf{ALG}$, 
the \textit{tightness gap}, denoted by $\kappa(\textsf{ALG})$, is defined as the smallest ratio between $\underline{T}(\textsf{ALG};\mathbf{p})$ and $\underline{T}^*(\mathbf{p})$, i.e.,
\begin{equation}
\label{eq:kappa}
\inf\left\{\kappa\in [1,\infty):\sup_{\mathbf{p}\in \mathcal{P}}\overline{T}(ALG;\mathbf{p}) \leq \kappa \sup_{\mathbf{p}\in \mathcal{P}} \underline{T}^*(\mathbf{p})\right\}.
\end{equation}
\end{definition}

This tightness gap highlights the optimality of the considered group testing algorithm, which measures how closely an algorithm's performance aligns with its theoretical optimal. This gap is crucial not only in presenting our main result regarding the consistency and robustness trade-off but also in applications such as large-scale energy systems, where near-optimal performance grants significant operational savings. A typical example related to our algorithm design is the Laminar Algorithm (\laalg) in~\cite{priorstats}, whose 
 \textit{tightness gap} $\kappa_{LA}$ is less than $2 + {6}/{\log(n/d)}$. In the sequel, we introduce a key performance benchmark for our novel probabilistic group testing model with ML predictions, by considering adaptive group testing as an instance of online optimization. 

\begin{definition}[Expected $\kappa$-Regret]
The \textit{expected $\kappa$-regret}, denoted by $\mathsf{ER}(\textsf{ALG};\varepsilon)$ with $D_{\mathrm{KL}}\left(p,\widetilde{p}\right)\leq \varepsilon$ for a given deterministic group testing algorithm $\textsf{ALG}$ is defined as
\begin{align}
\label{eq:er}
    \mathsf{ER}(\textsf{ALG};\varepsilon) \coloneqq \sup_{\mathbf{p}, \mathbf{\widetilde{P}}\in\mathcal{P}(\varepsilon)} \mathbb{E}\left[{T}(\textsf{ALG};\mathbf{\widetilde{p}}) - \kappa(\textsf{LA})  \underline{T}^*(\mathbf{p}) \right]
\end{align}
where $\mathcal{P}(\varepsilon) \coloneqq \left\{(p,q)\in \mathcal{P}\times\mathcal{P}: D_{\mathrm{KL}}\left(p,q\right) \leq \varepsilon\right\}$; $\kappa(\textsf{LA})$ denotes the tightness bound of the \laalg in~\cite{priorstats}; and the expectation is taken over randomly sampled instances of $\mathbf{X}$ distributed according to $\mathbf{p}$. 
\end{definition}

The notion of regret defined above characterizes the performance of $\textsf{ALG}$ by comparing the \textit{worst-case} difference between the expected number of tests required by $\textsf{ALG}$, given untrusted ML predictions $\widetilde{\mathbf{p}}$ that may contain some error, and a fundamental limit on the minimal expected number of tests required, normalized by the tightness gap.

\section{Detecting Malicious Users Efficiently from Behavioral Data} \label{sec:GT+MLPred}
In this section, we provide a detailed implementation of \ouralg, which is designed to equip with both the advantages of the probabilistic group testing with (untrusted) ML predictions and the classical group testing for worst-case situations and achieves near-optimal performance. First, \textsc{Procedure}~\ref{alg:PGT} summarizes the Laminar Algorithm (\laalg) in~\cite{priorstats}, which is an adaptive group testing algorithm that only works well with accurate prior statistics, i.e., the unknown exact $\mathbf{p}$ in our model. Second, \textsc{Procedure}~\ref{alg:CGT} summarizes the Generalized Binary Splitting (\gbs) algorithm~\cite{gbsa}, as a classic adaptive combinatorial group testing algorithm that does not incorporate any ML predictions. Therefore, even if the learned ML advice from the EV drivers' behavioral data is wrong, \textsc{Procedure}~\ref{alg:CGT} remains unaffected, while \textsc{Procedure}~\ref{alg:PGT} is sensitive to the error induced by the ML model. Both \textsc{Procedure}~\ref{alg:PGT} and \textsc{Procedure}~\ref{alg:CGT} are provided in Appendix~\ref{app:procedure}. Later in Section~\ref{sec:numexperiment} and Appendix~\ref{sec:simulatedexp}, we compare both \laalg and \gbs with our proposed robust detection algorithm.

\subsection{Warm-Up: Why Don't Trust All Predictions?}

Prior to delving into the \ouralg algorithm, it is crucial to understand the impact of untrusted ML predictions on established probabilistic group testing algorithms. A notable example is the \laalg algorithm detailed in~\cite{priorstats}. Theorem 1 from~\cite{priorstats} presents an upper bound on the expected number of tests required by \laalg, denoted as $\mathbb{E}[T(\laalg)]$, which is bounded from above by $2 \mathbb{H}(\mathbf{X}) + 6d$, with $\mathbb{H}(\mathbf{X})$ representing the binary entropy of $\mathbf{X}$. Building upon this, we introduce a lemma that expands the theorem to more practical scenarios involving ML predictions $\widetilde{\mathbf{p}}$, and offers a generalized upper bound on $\mathbb{E}[T(\laalg)]$ accordingly.

\begin{lemma}[Generalization of Theorem 1 in~\cite{priorstats}]
\label{lemma:generalization}
    When trusting the ML predictions $\widetilde{\mathbf{p}}$, the expected number of tests used by the \laalg satisfies
    \begin{align}
\label{eq:lemma_generalization_1}
\mathbb{E}\left[T(LA)\right] \leq 2\sum_{i=1}^{n} p_i \log \frac{1}{\widetilde{p}_i} + 6 d.
    \end{align}
\end{lemma}
The proof of Lemma~\ref{lemma:generalization} follows the same argument of the proof of Theorem 1 in~\cite{priorstats}, where the term $p_i$ in the summation of~\eqref{eq:lemma_generalization_1} follows by taking expectation according the true distribution of $X$ and the term $\widetilde{p}_i$ appears by constructing testing sub-trees based on the ML input $\widetilde{\mathbf{p}}$. A key consequence of Lemma~\ref{lemma:generalization} is observed when the predicted probabilities $\widetilde{p}_i$ are significantly low. Under such circumstances, the \laalg algorithm tends to combine users corresponding to these negligible probabilities with a large set of users. This can substantially degrade the \laalg's efficiency, thereby the upper bound in~\eqref{eq:lemma_generalization_1} is not tight. Such a scenario highlights the need for a design that ensures robustness even when confronted with untrustworthy ML predictions.

\subsection{Our Solution: Incorporating a Safety Threshold}

To mitigate the vulnerability issues raised by small $\widetilde{p}_i$'s, we now introduce a safety-enabled testing scheme, \ouralg, which combines both probabilistic and classical group testing methods. We consider a \textit{safety threshold}, denoted $\eta\geq \frac{1}{n}$, which divides the users $\mathcal{N}$ into two disjoint pools $\mathcal{N}_p(\eta)$ and $\mathcal{N}_c(\eta)$ such that $\mathcal{N}_p(\eta)\cup\mathcal{N}_c(\eta) = \mathcal{N}$, based on the associated ML predictions $\widetilde{\mathbf{p}}$. For the subset of users within $\mathcal{N}_p(\eta)$, the predicted probabilities $(\widetilde{p}_i : i \in \mathcal{N}_p(\eta))$ are considered reliable and are therefore utilized accordingly. Conversely, for users in the set $\mathcal{N}_c(\eta)$, we disregard the less reliable predictions and instead employ combinatorial group testing techniques to identify any potentially malicious behavior. The two disjoint user pools are defined formally in the following:
\begin{align}
\label{eq:p1}
    \mathcal{N}_p(\eta) \coloneqq &\left\{i\in\mathcal{N}:\widetilde{p}_i\geq \eta \right\},\\
\label{eq:p2}
    \mathcal{N}_c(\eta) \coloneqq  & \mathcal{N}\backslash \mathcal{N}_p(\eta).
\end{align}

\begin{figure}[!t]
    \removelatexerror
    \floatname{algorithm}{Algorithm}
    \begin{algorithm}[H]
    \caption{\setlength{\spaceskip}{0.2em plus 0.1em minus 0.1em} \textsf{G}roup \textsf{T}esting with \textsf{U}ntrusted ML \textsf{A}dvice (\ouralg)}
    \begin{algorithmic}[1]\label{alg:pseudocode}
    \STATE \textbf{Input:} Population set $\mathcal{N}$, distributional advice $\widetilde{\mathbf{p}}$
    \STATE Initialize a safety threshold $\eta\geq \frac{1}{n}$
    \STATE Partition $\mathcal{N}$ into $ \mathcal{N}_p(\eta)$ and $ \mathcal{N}_c(\eta)$ based on $\eta$ and $\widetilde{\mathbf{p}}$ according to~\eqref{eq:p1} and~\eqref{eq:p2}
    \FOR{$i$ in $\mathcal{N}_p(\eta)$}
    \STATE Implement \textsc{Procedure}~\ref{alg:PGT} using the predicted $\widetilde{p}$ on $\mathcal{N}_p(\eta)$
    \ENDFOR
    \FOR{$i$ in $\mathcal{N}_c(\eta)$}
        \STATE Implement \textsc{Procedure}~\ref{alg:CGT} on $\mathcal{N}_c(\eta)$ 
\ENDFOR
    \end{algorithmic}
    \end{algorithm}
\end{figure}

Enabled by a tunable safety threshold, the framework summarized in Algorithm~\ref{alg:pseudocode}, demonstrates a generic approach that integrates any recognized probabilistic group testing algorithm with predictions for users in $\mathcal{N}_p(\eta)$, as well as combinatorial group testing algorithms for users in $\mathcal{N}_c(\eta)$. In the sequel, we provide theoretical guarantees of \ouralg, assuming \laalg and \gbs are implemented as inner procedures in Algorithm~\ref{alg:pseudocode}. It is worth noting that the algorithm we present functions as a meta-testing scheme. It is designed to integrate and accommodate a variety of testing procedures -- a probabilistic one that ensures \textit{consistency} when prediction errors are low, and a combinatorial one that guarantees \textit{robustness} against increasing prediction inaccuracies. The procedures can be arbitrary, as long as each is selected in its capacity as satisfying one of these crucial properties, allowing the overall scheme to maintain performance across varying levels of prediction reliability. In \textsc{Procedure} \ref{alg:PGT} and \textsc{Procedure} \ref{alg:CGT} we exemplify the \laalg and the Generalized Binary Splitting (\textsf{GBS}) as two particular instances, provided in Appendix~\ref{app:procedure}. 

\section{Theoretical Results} \label{sec:theories}
In this section, we provide theoretical analysis on \ouralg and show that it achieves near-optimal performance, in terms of the consistency and robustness trade-off (see Definition~\ref{def:tradeoff}). Fixing the inner procedures as \laalg and \gbs, respectively, we bound the expected regret of \ouralg, with a tightness gap $\kappa(\laalg)$.

\subsection{Expected Regret Analysis}

\begin{theorem}[Expected $\kappa(\laalg)$-Regret] \label{thm:regret}
Fix $D_{\mathrm{KL}}\left(p,\widetilde{p}\right)\leq\varepsilon$, when $n\geq d\geq e$. The expected $\kappa(\laalg)$-regret of \ouralg satisfies
\begin{align*}
    \mathsf{ER}\left(\ouralg~;\varepsilon\right) &\leq 2\min \big\{d \log (1/\eta), \varepsilon_p\big\}\\
    &+ 2\min \big\{d \log n , \varepsilon_c + d \log \eta n \big\},
\end{align*}
where $\varepsilon_p$ and $\varepsilon_c$ are upper bounds on the pseudo KL-divergence for the two pools, $\mathcal{N}_p(\eta)$ and $\mathcal{N}_c(\eta)$ respectively, such that $\sum_{i\in \mathcal{N}_p(\eta)} \left(p_i\log {p_i}/{\widetilde{p}_i} \right)\leq \varepsilon_p$ and $\sum_{i\in \mathcal{N}_c(\eta)} \left(p_i\log {p_i}/{\widetilde{p}_i} \right)\leq \varepsilon_c$.
\end{theorem}

To characterize the overall performance of \ouralg, following the standard definitions in the online algorithms community~\cite{wei2020optimal,aldridge2019group}, we adapt the following definitions to evaluate the consistency and robustness of the algorithm with respect to two error regimes of the expected regret.

\begin{definition}[Consistency and robustness]
    \label{def:tradeoff}
A testing algorithm $\mathsf{ALG}$ is $\gamma$-consistent if its expected $\kappa(\laalg)$-regret satisfies $\mathsf{ER}(\mathsf{ALG};\varepsilon)\leq \gamma$ when $\varepsilon = 0$, and $\beta$-robust if  $\mathsf{ER}(\mathsf{ALG};\varepsilon)\leq \beta$ for any $\varepsilon\geq 0$.
\end{definition}
Here, the values of $\gamma$ and $\beta$ depend on system parameters such as $d$ and $n$. Theorem~\ref{thm:regret} above implies the following result.
\begin{corollary}
\label{coro:tradeoff}
 With a fixed threshold parameter $\eta\geq \frac{1}{n}$, the \ouralg algorithm  is $\left(d \log \left(\eta n\right)\right)$-consistent and $\left(4d\log\frac{n}{\eta}\right)$-robust.
\end{corollary}
This result highlights that when $\eta=\Theta(1/n)$, the proposed \ouralg is $O(d)$-consistent, implying that it uses at most $O(T(\laalg))$ number of tests, which is optimal when the ML predictions $\widetilde{\mathbf{p}}$ are perfect. On the other hand, for any $\varepsilon$, \ouralg is $O(d\log n)$-robust. Note that any group testing algorithm must have a lower bound $\Omega(d\log n)$ on its expected number of tests, as implied by the information-theoretic lower bound~\cite{du2000combinatorial,comp}.
Furthermore, the following result verifies that the consistency and robustness trade-off achieved by \ouralg in Corollary~\ref{coro:tradeoff} is near-optimal.

\begin{theorem} 
\label{fundamental limits}
    For any deterministic group testing algorithm $\mathsf{ALG}$, if it is $\left(\frac{1}{2} d \log (n \eta)\right)$-consistent, then it is at least $O\left (d \log \frac{n}{\eta} \right )$-robust.
\end{theorem}
Note that \ouralg does not use any randomness in the algorithm implementation; it is therefore a deterministic algorithm.

\section{Best-of-Both-Worlds Performance of \ouralg} \label{sec:numexperiment}
In this part, we will use synthetic data to test the performance of the aforementioned three group testing algorithms against various values of error (in terms of the KL divergence between $p$ and $\widetilde{p}$, denoted by $D_{\mathrm{KL}}\left(p,\widetilde{p}\right)$) to verify the efficacy of \ouralg and the robustness of its performance, regardless of the prediction error of the ML model. The detailed setting is deferred to Appendix~\ref{sec:numexperiment}. A simulated V2G system used to validate the applicability and effectiveness of \ouralg in real-life scenarios is deferred to Appendix~\ref{sec:simulatedexp}.
\begin{figure}[!h]
    \centering
\includegraphics[width = 0.7\textwidth]{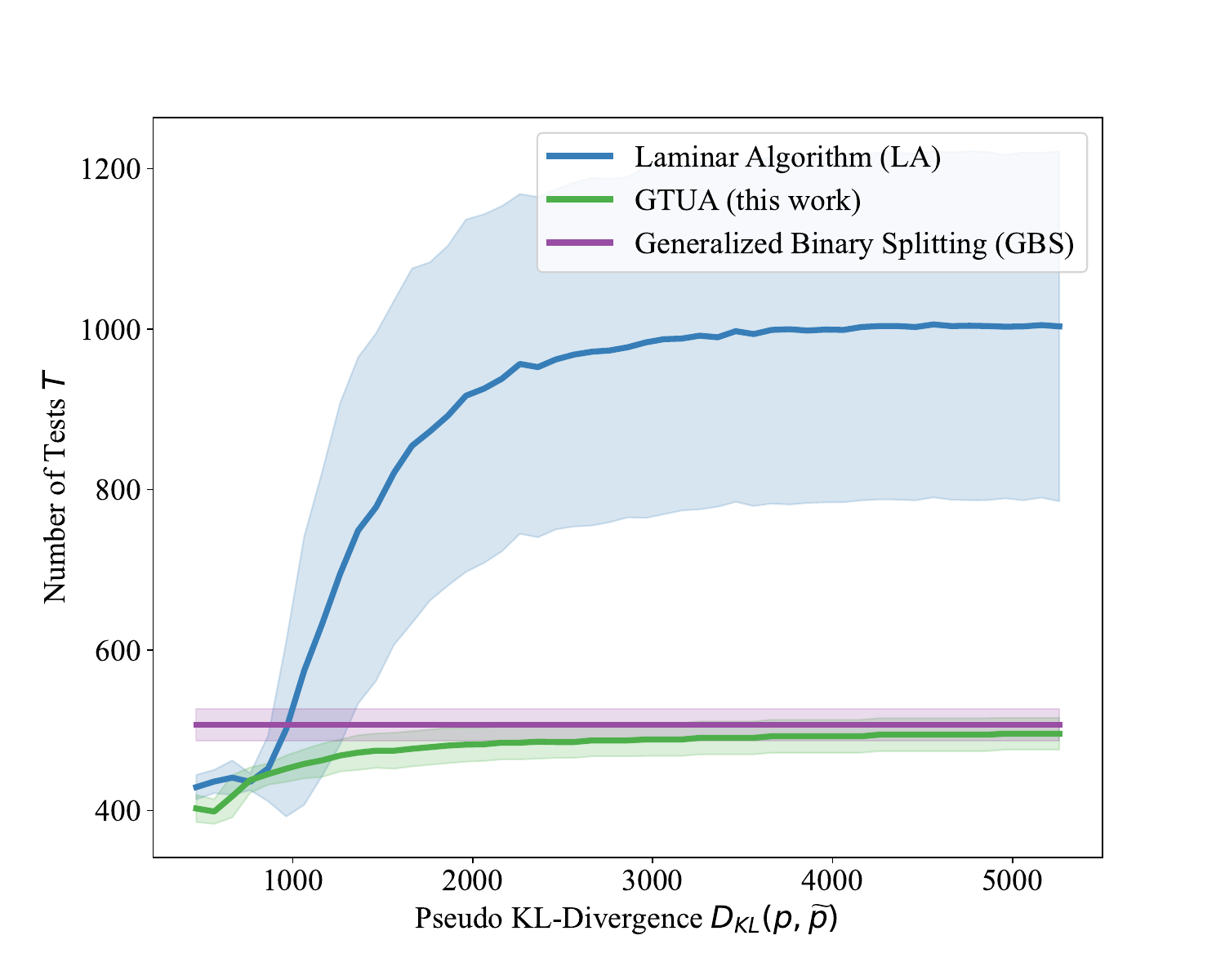}
\caption{\textbf{Comparison of the average number of tests} for \laalg~\cite{priorstats}, \gbs~\cite{gbsa}, and the proposed \ouralg (Algorithm~\ref{alg:pseudocode}) in this work. Shadow areas depict the magnitudes of standard deviations.}
    \label{fig:test}
\end{figure}

\noindent \textbf{Experimental Results.} The result in Figure~\ref{fig:test}  highlights that when the divergence between \( \mathbf{p} \) and \(\widetilde{\mathbf{p}} \) is small (implying high predictive accuracy), the \laalg~\cite{priorstats} exhibits high efficiency as desired. However, when prediction error increases, its efficacy degrades significantly, as evidenced by the escalating number of required tests and high variance in the figure (see the blue curve). In contrast, \gbs~\cite{gbsa}'s performance (see the purple curve) remains unaffected regardless of the prediction error of the distributional advice $\widetilde{\mathbf{p}}$.  As suggested by our theoretical guarantees in Theorem~\ref{thm:regret} and Corollary~\ref{coro:tradeoff}, the proposed algorithm shows its reliability in practical scenarios, where ML predictions cannot be fully trusted.

\section{Conclusion} \label{sec:conclude}
In this paper, we addressed the critical issue of detecting malicious EV drivers in vehicle-to-grid systems using group testing with untrusted ML predictions. We proposed \ouralg, a safety-enhanced group testing scheme that combines the efficiency of probabilistic group testing with ML predictions and the robustness of combinatorial group testing. Theoretically, we proved that \ouralg achieves a near-optimal consistency and robustness trade-off. We validated the effectiveness of \ouralg through experiments on synthetic data and simulations on real-world EV charging data, demonstrating its superior performance compared to existing group testing algorithms.

Built upon the results presented in this work, there are various aspects that are worth future investigation. For example, it would be interesting to study if more advanced ML models can further improve the detection efficiency in practice, even though our algorithm has a theoretically guaranteed performance regardless of the ML prediction accuracy. 
The proposed \ouralg algorithm offers a promising solution to adaptively detect malicious EVs in a V2G system with learned distributional advice from the EVs' behavioral data. Extending \ouralg to non-adaptive settings, where all tests are designed offline beforehand, is a potential direction for future research.

\bibliographystyle{unsrt}  
\bibliography{references}

\onecolumn{
\appendices
\section{Probabilistic and Combinatorial Group Testing Procedures}
\label{app:procedure}
\begin{figure}[!h]
    \removelatexerror
    \begin{algorithm}[H]
    \caption{\textsc{Procedure} \textsf{PGT}~\cite{priorstats}}
    \begin{algorithmic}[1]\label{alg:PGT}
    \Procedure{PGT}{$\mathcal{N}_p\left(\eta\right), \widetilde{\mathbf{p}}$} $\triangleright$\textcolor{gray}{\textit{This procedure implements the Maximum Entropy-based \laalg~\cite{priorstats} for group testing}}
    \BeginProcedure
        \STATE \textbf{Laminar Algorithm (LA):}\\
         \textsf{First Stage}
         \STATE Partition $\mathcal{N}$ into subsets $\{S^r_{1,1}\}_{r=1}^{m}$ greedily:$\triangleright$ \textcolor{gray}{\textit{$S_{k, l}^{r}$(k, l are parameters indexing the child nodes and r indexes the
partitions)}}
\begin{equation*}
    \min_{S_{1,1}^{r}} \left| \prod_{X_i \in S_{1,1}^{r}} (1 - p_i) - \frac{1}{2} \right|,
\end{equation*}
\begin{equation*}
    \text{subject to } S_{1,1}^{r} \subseteq \mathcal{N} \setminus \bigcup_{j=1}^{r-1} S_{1,1}^{r}.
\end{equation*}
        \textsf{The $k$-th Stage ($k \in \{2, 3, 4,  \ldots\}$)}
        \FOR{subgroups that tested positive in the previous stage}
      \STATE Partition the set $S^r_{k-1, l}$ according to:
      \begin{equation*}
          \min_{S_{k,l}^{r}} \left| \frac{1 - \prod_{X_i \in S_{k,l}^{r}} (1 - p_i)}{1 - \prod_{X_i \in S_{k-1,l}^{r}} (1 - p_i)} - \frac{1}{2} \right|
,
      \end{equation*}
\begin{equation*}
    \text{subject to } S_{k,l}^{r} \subseteq S_{k-1,l}^{r}.
\end{equation*}
    \ENDFOR
\STATE This strategy is designed to maximize the entropy, which is optimal for probabilistic group testing problem.
\EndProcedure
    \end{algorithmic}
    \end{algorithm}
\end{figure}

\begin{figure}[!h]
    \removelatexerror
    \begin{algorithm}[H]
    \caption{\textsc{Procedure} \textsf{CGT}~\cite{gbsa}}
    \begin{algorithmic}[1]\label{alg:CGT}
    \Procedure{CGT}{$\mathcal{N}_c(\eta)$}
$\triangleright$\textcolor{gray}{\textit{This procedure implements the \gbs~\cite{gbsa} for group testing}}
\BeginProcedure
\STATE \textbf{Generalised Binary-Splitting (GBS) Key Steps:}
\STATE Set \( \alpha = \lceil \log_2 \frac{n - d + 1}{d} \rceil \).
\STATE Perform group testing with groups of size \( 2^\alpha \).

\IF{a group is tested positive}
\STATE Use binary search to find a defective.
\ENDIF
\STATE Adjust \( n \) and \( d \) after identifying defectives and repeat.
\EndProcedure
    \end{algorithmic}
    \end{algorithm}
\end{figure}

\section{Useful Lemmas}
In this appendix, we provide lemmas that will be used in the proofs of our main results.

\begin{lemma}[Theorem 1 in~\cite{priorstats}] For any group testing algorithm operating on the user set $\mathcal{N}$ with the probability vector $\mathcal{P}$, and allowing a maximum error probability $P_e$, the lower bound on the number of tests is given by:
\begin{equation}
\label{lower bound}
    T \geq (1 - \mathbb{P}_e) H(\mathbf{X}),
\end{equation}
where $H(\mathbf{X}) \coloneqq \left(\sum_{i \in \mathcal{N}} p_i \log \frac{1}{p_i} + \sum_{i \in \mathcal{N}} (1 - p_i) \log \frac{1}{1 - p_i}\right)$ denotes the summation of entropy of random variable $X_i \in \mathbf{X}$.\\

The next lemma provides the upper bound of the expected number of tests required by \laalg~\cite{priorstats} (see Procedure~\ref{alg:PGT}).
\end{lemma}

\begin{lemma}[Theorem 2 in~\cite{priorstats}]
\label{lemma:priorstats}
A user $i$ with probability $p_i$ to be malicious needs at most \(\log \frac{1}{p_i}\) number of tests, and the expected number of tests $\mathbb{E}(T(\laalg))$ for the \textit{Laminar Algorithm} with a user set $\mathcal{N}$ and ground truth probability vector $\mathbf{p}$ can be bounded as:
    \begin{align}
    \nonumber
\mathbb{E}\left[T(\laalg)\right] & \leq 2\sum_{i \in \mathcal{N}} p_i \left\lceil \log\left(\frac{1}{p_i}\right) \right\rceil + 4\sum_{i \in \mathcal{N}} p_i \\
\nonumber
&\leq 2\sum_{i \in \mathcal{N}} p_i \log\left(\frac{1}{p_i}\right)  + 6\sum_{i \in \mathcal{N}} p_i.
    \end{align}

In the next lemma, we characterize the \textit{tightness bound} for the \textit{Laminar Algorithm} (\laalg), as a critical benchmark in probabilistic group testing.
\end{lemma}

\begin{lemma}
The \textit{tightness gap} $\kappa(\laalg)$ for \textit{(adaptive) Laminar Algorithm} (\laalg) in ~\cite{priorstats} is at most $2 + {6}/{\log(n/d)}$.
\begin{proof}
By definition, $\kappa(\laalg)$ can be computed as:
    \begin{align}
    \nonumber
        \kappa &= \lim_{n \to \infty}\frac{\max_{\mathcal{P}} \left ( 2\sum_{i \in \mathcal{N}} p_i \log(\frac{1}{p_i}) + 6\sum_{i \in \mathcal{N}} p_i \right )}{\max_{\mathcal{P}} (1 - \mathbb{P}_e) H(\mathbf{X})}\\
        \nonumber
        & \leq \lim_{n \to \infty}\frac{\max_{\mathcal{P}} \left ( 2\sum_{i \in \mathcal{N}} p_i \log(\frac{1}{p_i}) + 6\sum_{i \in \mathcal{N}} p_i \right )} {\max_{\mathcal{P}} (1 - \mathbb{P}_e) (\sum_{i \in \mathcal{N}} p_i \log(\frac{1}{p_i}))} \label{vanishingprob}\\
        \nonumber
        & = \frac{2 d \log(\frac{n}{d}) + 6d}{d \log(\frac{n}{d})}\\
        & = 2 + \frac{6}{\log(n/d)},
    \end{align}
where \eqref{vanishingprob} follows that in probabilistic group testing, $\mathbb{P}_e \to 0$ when $n \to \infty$. Furthermore, since $d = o(n)$, when $n \to \infty$, $\kappa \to 2$.
\end{proof}
\end{lemma}

Next, we consider the number of tests required by the generalized binary splitting (\gbs) algorithm in~\cite{gbsa}.
\begin{lemma}[Theorem 1 in~\cite{gbsa}]
\label{lemma:gbs}
Consider a population of $n$ users among which $d$ users are malicious, and suppose $d = o(n)$. In this context, the required number of tests of \gbs, denoted by $T(\gbs)$, can be bounded from above as follows:
\begin{equation}\label{testnumberG2}
    T(\gbs) \leq d\left(\lfloor \log \frac{n-d+1}{d} \rfloor + 2\right) + p - 1,
\end{equation}
where $p$ is a non-negative integer satisfying $0 \leq p < d$. 
\end{lemma}

Therefore, the number of tests $T(\gbs)$ satisfies $T(\gbs)\leq 2d\log n$ for $n\geq d\geq e$.

\section{Proof of Theorem~\ref{thm:regret}}
\begin{proof}[Proof of Theorem \ref{thm:regret}]Applying lemma~\ref{lemma:priorstats}, the expected number of tests for probabilistic group testing algorithm $\mathbb{E}[T_{i \in \mathcal{N}_p(\eta)}]$ can be bounded as:
\begin{align}
\nonumber
\mathbb{E}\left[T_{i \in \mathcal{N}_p(\eta)}\right] &\leq 2\sum_{i \in \mathcal{N}} p_i \lceil \log(\frac{1}{\widetilde{p}_i}) \rceil + 4\sum_{i \in \mathcal{N}} p_i\\
  &\leq 2\sum_{i \in \mathcal{N}} p_i \log(\frac{1}{\widetilde{p}_i}) + 2\sum_{i \in \mathcal{N}} \widetilde{p}_i + 4\sum_{i \in \mathcal{N}} p_i\label{testnumberG1}.
\end{align}

Therefore, combining \eqref{testnumberG1}, \eqref{testnumberG2}, \eqref{lower bound}, and using Lemma~\ref{lemma:generalization}, the expected $\kappa(\laalg)$-regret (see Definition~\ref{def:regret}) can be bounded as:
\begin{equation}\label{eq:er_1}
     \mathsf{ER}(\ouralg) \leq 2 \underbrace{\left (\sum_{i \in \mathcal{N}_p(\eta)}p_i \log \frac{1}{\widetilde{p}_i} + \sum_{i \in \mathcal{N}_p(\eta)}p_i + \sum_{i \in \mathcal{N}_p(\eta)}2\widetilde{p}_i \right )}_{\text{Lemma~\ref{lemma:generalization}}} 
    + \underbrace{\left(\sum_{i \in \mathcal{N}_c(\eta)} 3p_i \log\left(\frac{\vert \mathcal{N}_c(\eta) \vert}{\sum_{i \in \mathcal{N}_c(\eta)} p_i}\right) \right)}_{\text{Lemma~\ref{lemma:gbs}}} - \kappa H(\mathbf{X}).
\end{equation}
Rearranging the terms in~\eqref{eq:er_1} above and applying Lemma~\ref{lemma:gbs}, the expected regret satisfies
\begin{align}
\nonumber
    \mathsf{ER}(\ouralg) & \leq  \sum_{i \in \mathcal{N}_p(\eta)} 2p_i \log \frac{1}{\widetilde{p}_i} + \sum_{i \in \mathcal{N}_c(\eta)} 2p_i \log(\vert \mathcal{N}_c(\eta) \vert) - \sum_{i \in \mathcal{N}} 2p_i \log \frac{1}{p_i}\\
    \nonumber
    & \leq \sum_{i \in \mathcal{N}_p(\eta)}2p_i \log \frac{p_i}{\widetilde{p}_i} + \sum_{i \in \mathcal{N}_c(\eta)} 2 p_i \log(p_i \vert \mathcal{N}_c(\eta) \vert)\\
     &= \sum_{i \in \mathcal{N}_p(\eta)} 2p_i \log \frac{p_i}{\widetilde{p}_i} + \sum_{i \in \mathcal{N}_c(\eta)} 2 p_i \log \frac{p_i}{\widetilde{p}_i}
\label{eq:app_2}
     + \sum_{i \in \mathcal{N}_c(\eta)} 2p_i \log\left(\widetilde{p}_i \vert \mathcal{N}_c(\eta) \vert\right)
\end{align}
Since the threshold parameter $\eta$ partitions the population set $\mathcal{N}$ based on the rules in \eqref{eq:p1}-\eqref{eq:p2},~\eqref{eq:app_2} yields
\begin{align*}
    \mathsf{ER}(\ouralg)   & \leq 2\min \left\{\sum_{i \in \mathcal{N}_p(\eta)}p_i \log \frac{1}{\eta}, \varepsilon_p\right\}  + 2\min \left\{\sum_{i \in \mathcal{N}_c(\eta)}p_i \log n , \varepsilon_c + \sum_{i \in \mathcal{N}_c(\eta)} \log \eta \vert \mathcal{N}_c(\eta) \vert\right\}\\
    & \leq 2\min \left\{d \log \frac{1}{\eta}, \varepsilon_p\right\} + \min \{d \log n , \varepsilon_c + d \log \eta n\}
\end{align*}
where the terms $\varepsilon_p$ and $\varepsilon_c$ are defined in Theorem~\ref{thm:regret}.
\end{proof}

\section{Proof of Theorem~\ref{fundamental limits}}
\begin{proof}[Proof of Theorem~\ref{fundamental limits}]Suppose that a probabilistic group testing algorithm ALG is $\frac{1}{2} \log n \lambda$-consistent. Assuming that the true probability vector $\mathbf{p}$ and the predicted probability vector $\widetilde{\mathbf{p}}$ are the same, then we need at least $d$ tests to find out all the defective cases. Therefore, the algorithm ALG needs at least $d + \frac{1}{2} \log n \lambda$ number of tests to find out all defective individuals. Next, we assume that $\widetilde{\mathbf{p}}$ and $\mathbf{p}$ have totally different distributions. Thus, the robustness constant $\beta$ can be bounded as follows.
\begin{align}
\nonumber
   & \beta  \geq \log \left ( \frac{\tbinom {n-d} {d}}{2^{d + \frac{1}{2} \log n \lambda}} \right )
    = \log \left ( \frac{\tbinom {n-d} {d}}{(2 \sqrt{n \lambda})^d} \right )\\
   &=O\left( \log \left( \frac{\sqrt{2 p_i (n-d) }(\frac{n-d}{e})^{n-d}}{\sqrt{2 p_i (n-2d) }(\frac{n-2d}{e})^{n-2d} \sqrt{2 p_i d }(\frac{d}{e})^{d} (2 \sqrt{n \lambda})^d} \right) \right)\label{eq:Stirling Approx},
\end{align}
where we have applied the
 Stirling's approximation to derive~\eqref{eq:Stirling Approx}. Continuing from above,
\begin{align}
\label{eq:tradeoff_1}
  \beta  & \geq O\left(\log \left ( \frac{(n-d)^{n-d+\frac{1}{2}}}{\sqrt{2 p_i}d^{d + \frac{1}{2}}(n-2d)^{n-2d+\frac{1}{2}} (2 \sqrt{n \lambda})^d} \right )\right),
 \end{align} 
with the term inside $O(\cdot)$ in~\eqref{eq:tradeoff_1} bounded from above by
\begin{align*}
&\frac{(n-d)^{n-d+\frac{1}{2}}}{\sqrt{2 p_i}d^{d + \frac{1}{2}}(n-2d)^{n-2d+\frac{1}{2}} (2 \sqrt{n \lambda})^d}=  O\left(\log \frac{n}{\lambda}\right),
\end{align*}
implying the theorem.
\end{proof}

\section{Numerical Experiment Settings.} \label{sec:numexpsetting}
Recall all the parameters in \ouralg, $n$ denotes the total EV users under monitoring, $\mathbf{p}$ is the true probability vector for each EV user to be defective, and $\widetilde{\mathbf{p}}$ is the machine-learned prediction for $\mathbf{p}$. Setting $n = 1000$, we fix $\mathbf{p}$ and vary the distributional advice $\widetilde{\mathbf{p}}$, which results in a sequence of pseudo KL-divergence $D_{\mathrm{KL}}\left(p,\widetilde{p}\right)$ values ($x$-axis in Figure~\ref{fig:test}) defined in~\eqref{eq:sum_KL}.

\section{Case Studies of Real-World \textsc{ACN-Data}} \label{sec:simulatedexp}
In this section, we will use authentic EV driver behavior data to validate the effectiveness of the proposed algorithm. We will first explain how we learn EV driver behavior from \textsc{ACN-Data}~\cite{lee_acndata_2019}, which contains ``\textsf{userID}", ``\textsf{connectionTime}", ``\textsf{doneChargingTime}", ``\textsf{disconnectTime}", ``\textsf{requestedDeparture}". Then we will elaborate our simulation process. Finally, simulation results will be displayed to show that \ouralg can efficiently detect malicious EV drivers in a V2G system.

\paragraph{Learn EV driver behavior from real world EV charging dataset.}
Using the similar idea in~\cite{lee_acndata_2019}, we first utilize Gaussian Mixture Models (GMMs)~\cite{lindsay1995mixture} to approximate an distribution of EV drivers' behavior. The dataset, which includes the data from Jan. 1, 2019 to Sep. 1,2021, can be modelled into a GMM model based on the following assumption: for each EV driver, their behavior profiles is finite~\cite{lee_acndata_2019}. To be more specific, for each charging session \(i\) in the dataset, we denote the corresponding user profile using a triple \(x_i = (a_i, d_i, e_i)\) in \(\mathbb{R}^3\), where \(a_i\) stands for the arrival time, \(d_i\) denotes the duration (``\textsf{disconnectTime}" - ``\textsf{connectionTime}"), and \(e_i\) is the deviation ("\textsf{disconnectTime}" - "\textsf{requestedDeparture}"). Therefore, each data point \(x_i\) can be viewed as an user profile with a certain probability. The GMM model is trained on the whole data set, and we generate $100, 000$ samples from the trained GMM model to evaluate how well this model fits the underlying distribution. Figure~\ref{fig:gmmvalidation} shows that the trained GMM model matched the real world data well. 
\begin{figure}[!t]
    \centering
\includegraphics[width = 0.6\textwidth]{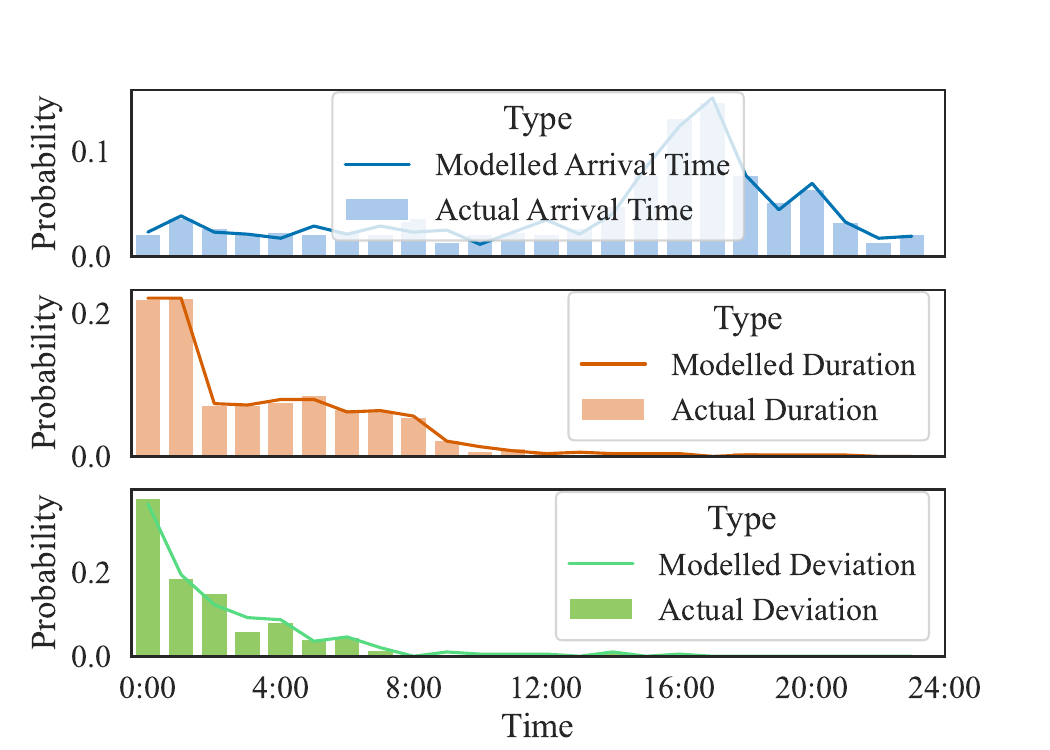}
\caption{Comparison of model distributions with actual collected data.}
    \label{fig:gmmvalidation}
\end{figure}
The trained GMM model is also used to predict whether an EV driver is likely to be malicious based on its prediction of the deviation. If the deviation is greater than $2$ hours, it indicates a higher likelihood of the user not adhering to the schedule, which represents the trustworthiness of the user. This metric can be used to infer whether the user will be malicious or non-cooperative in the V2G system. 

\paragraph{Simulation setting}
We first sample $100, 000$ copies of EV drivers' profile from the trained GMM model, with each profile including arrival time, duration, and deviation. The simulated V2G system starts with no EVs connected, and EVs are added to the system gradually based on their arrival time at each time step $t$. For those users with deviation greater than $2$ hours, we classify them as malicious users. These users may leave before the scheduled time, causing the total energy injection to be lower than the expectation. To ensure timely detection of malicious EV drivers, we conduct testing every hour. During the testing process, we implement \ouralg by disconnecting a subset of EV each time, which allows us to perform the grouping operations necessaty for identifying non-cooperative drivers.

\paragraph{Results}
The result in Figure~\ref{fig:finalresult} illustrates that the number of tests condected each hour is significantly lower than number of EVs being tested. Compared with individually testing all the EVs in the V2G system, \ouralg reduces the number of tests by about $40$ to $80$ percent. This observation further validate the efficiency of \ouralg in real world scenario. By minimizing the number of tests required to accurately identify malicious EV drivers, \ouralg reduced computational overhead. The reliable performance of \ouralg will enhance the overall effciency and reliability of the V2G system. In total we conducted five experiments using different random seeds, and the average results are shown in the table~\ref{tab:ratio_part2}.

\begin{figure}[!t]
    \centering
\includegraphics[width = 0.6\textwidth]{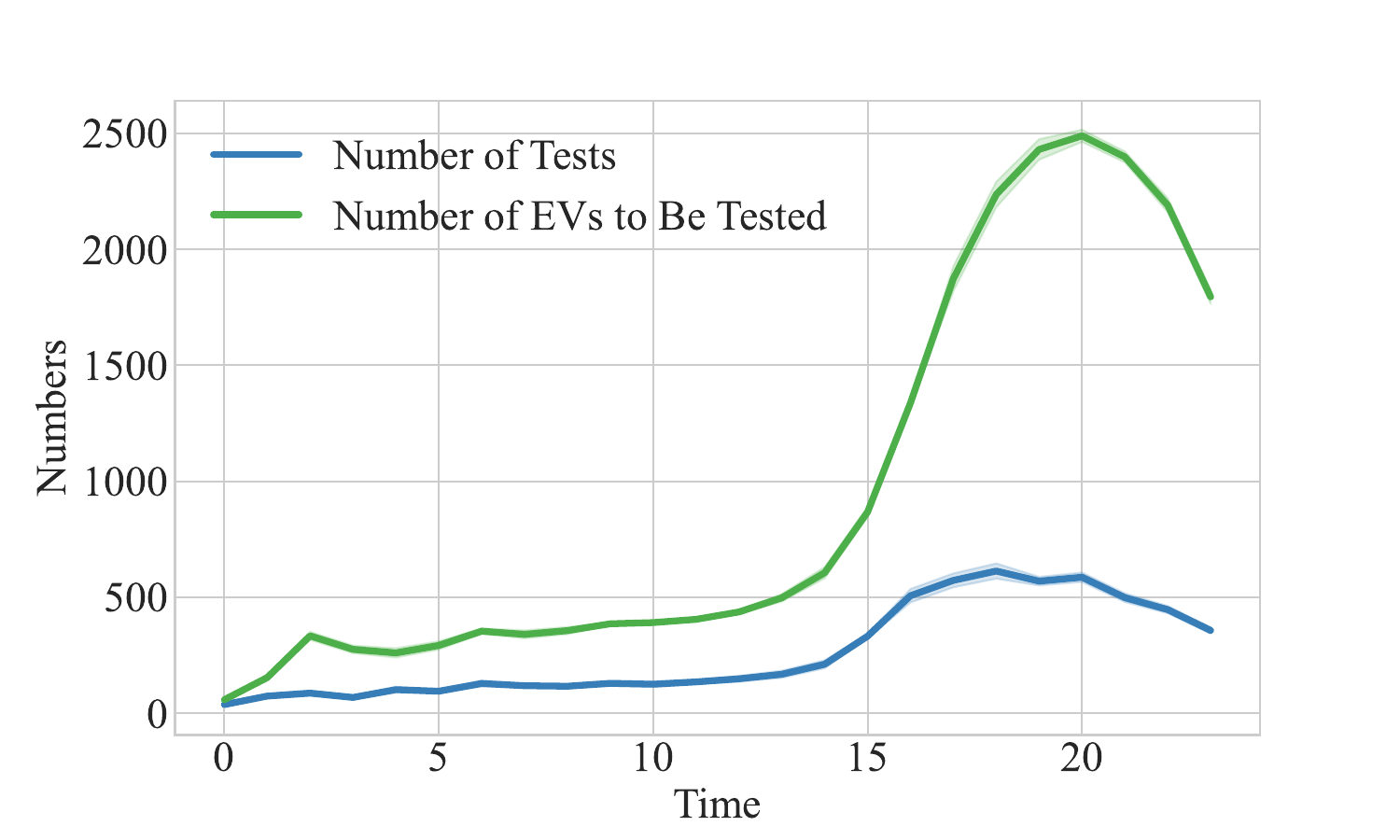}
\caption{Comparison of number of tests each hour using \ouralg and the number of EVs to be tested each hour.}
    \label{fig:finalresult}
\end{figure}

\begin{table}[!t]
\centering
\scalebox{0.9}{
\begin{tabular}{|c|*{12}{c|}}
\hline
 & 1 & 2 & 3 & 4 & 5 & 6 & 7 & 8 & 9 & 10 & 11 & 12 \\ \hline
Ratio & 0.64 & 0.48 & 0.26 & 0.25 & 0.39 & 0.32 & 0.36 & 0.35 & 0.33 & 0.33 & 0.32 & 0.33 \\  \hline
Number of Tests & 37.2 & 73.8 & 86.6 & 68.0 & 101.6 & 94.8 & 128.2 & 118.6 & 116.4 & 128.8 & 125.0 & 135.0 \\ \hline
Number of Users & 58.2 & 153.6 & 333.6 & 275.4 & 260.0 & 292.2 & 353.8 & 340.0 & 356.2 & 385.6 & 390.8 & 404.8 \\ \hline

\hline
 & 13 & 14 & 15 & 16 & 17 & 18 & 19 & 20 & 21 & 22 & 23 & 24 \\ \hline
Ratio & 0.34 & 0.34 & 0.35 & 0.38 & 0.38 & 0.31 & 0.27 & 0.23 & 0.24 & 0.21 & 0.20 & 0.20 \\  \hline
Number of Tests & 148.4 & 168.0 & 211.2 & 332.2 & 506.6 & 572.8 & 613.0 & 569.4 & 586.4 & 498.2 & 446.6 & 356.8 \\ \hline
Number of Users & 436.6 & 497.6 & 606.0 & 867.6 & 1337.8 & 1873.4 & 2236.6 & 2430.8 & 2489.6 & 2398.4 & 2190.0 & 1795.0 \\ \hline
\end{tabular}}
\caption{Number of tests each hour using \ouralg and the number of EVs to be tested each hour.}
\label{tab:ratio_part2}
\end{table}

}

\end{document}